\documentclass[aps,prb,floatfix,amsmath,amssymb,eqsecnum,nofootinbib,twocolumn]{revtex4}
\usepackage{graphicx}% Include figure files
\usepackage{dcolumn}% Align table columns on decimal point
\usepackage{bbm,bm}% bold math
\usepackage{epstopdf}
\usepackage[bookmarks=true,colorlinks,linkcolor=blue,urlcolor=blue,citecolor=blue]{hyperref}
\usepackage{color}

\usepackage{setspace} %line spacing
\newcommand{\beq}{\begin{equation}}
\newcommand{\eeq}{\end{equation}}
\newcommand{\beqn}{\begin{eqnarray}}
\newcommand{\eeqn}{\end{eqnarray}}

\usepackage{amsmath}
\usepackage{amssymb}

\def \v{{\mathbf{v}}}
\def \u{{\mathbf{u}}}
\def \q{{\mathbf{q}}}
\def \R{{\mathbf{R}}}
\def \Q{{\mathbf{Q}}}

\def \q{{\mathbf{q}}}

\def \u{{\mathbf{u}}}

\def \R{{\mathbf{R}}}

\def \S{{\mathbf{S}}}

\newcommand{\elemA}{\mathord{\vcenter{\hbox{\includegraphics[height=6ex]{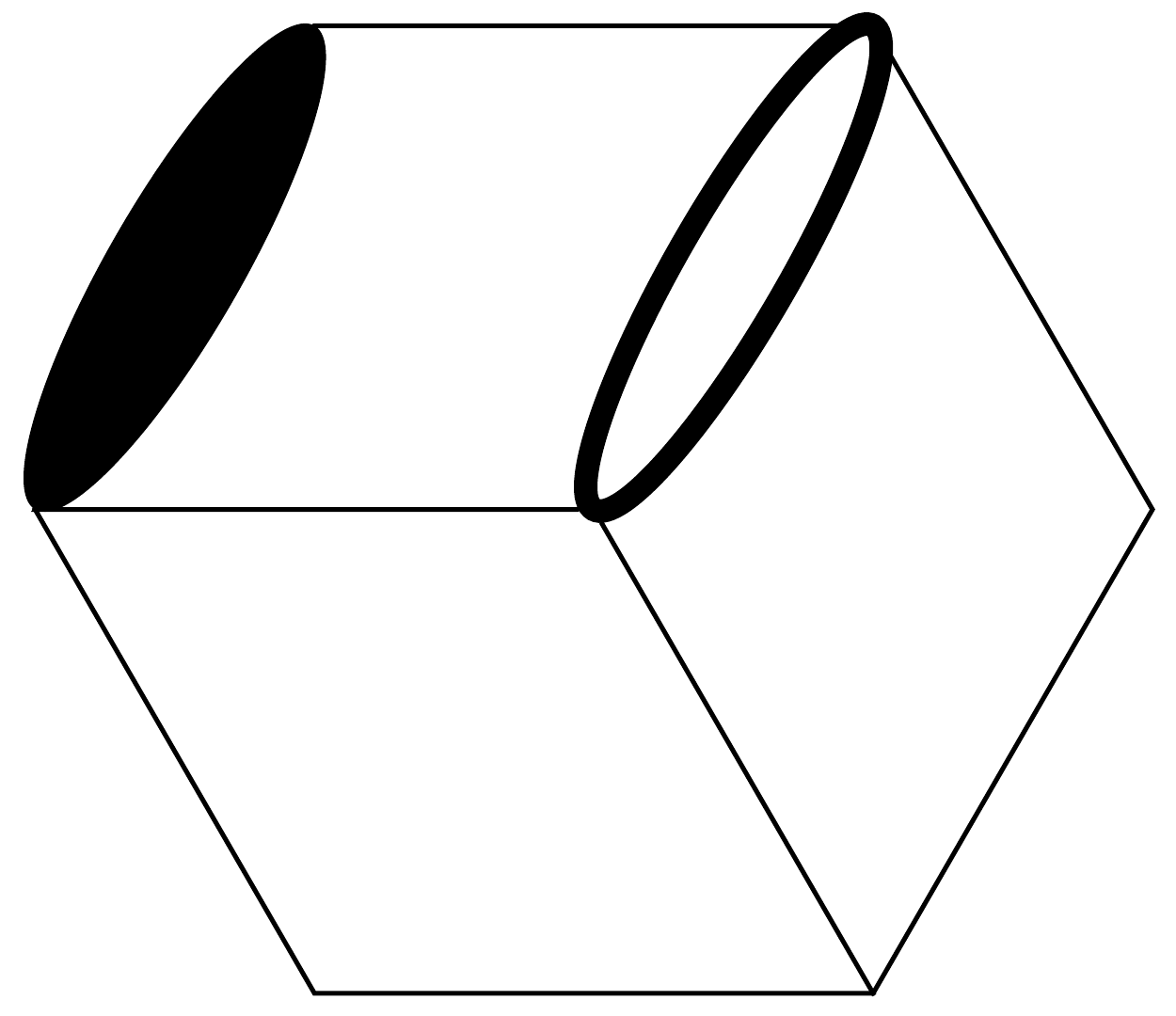}}}}}
\newcommand{\elemB}{\mathord{\vcenter{\hbox{\includegraphics[height=6ex]{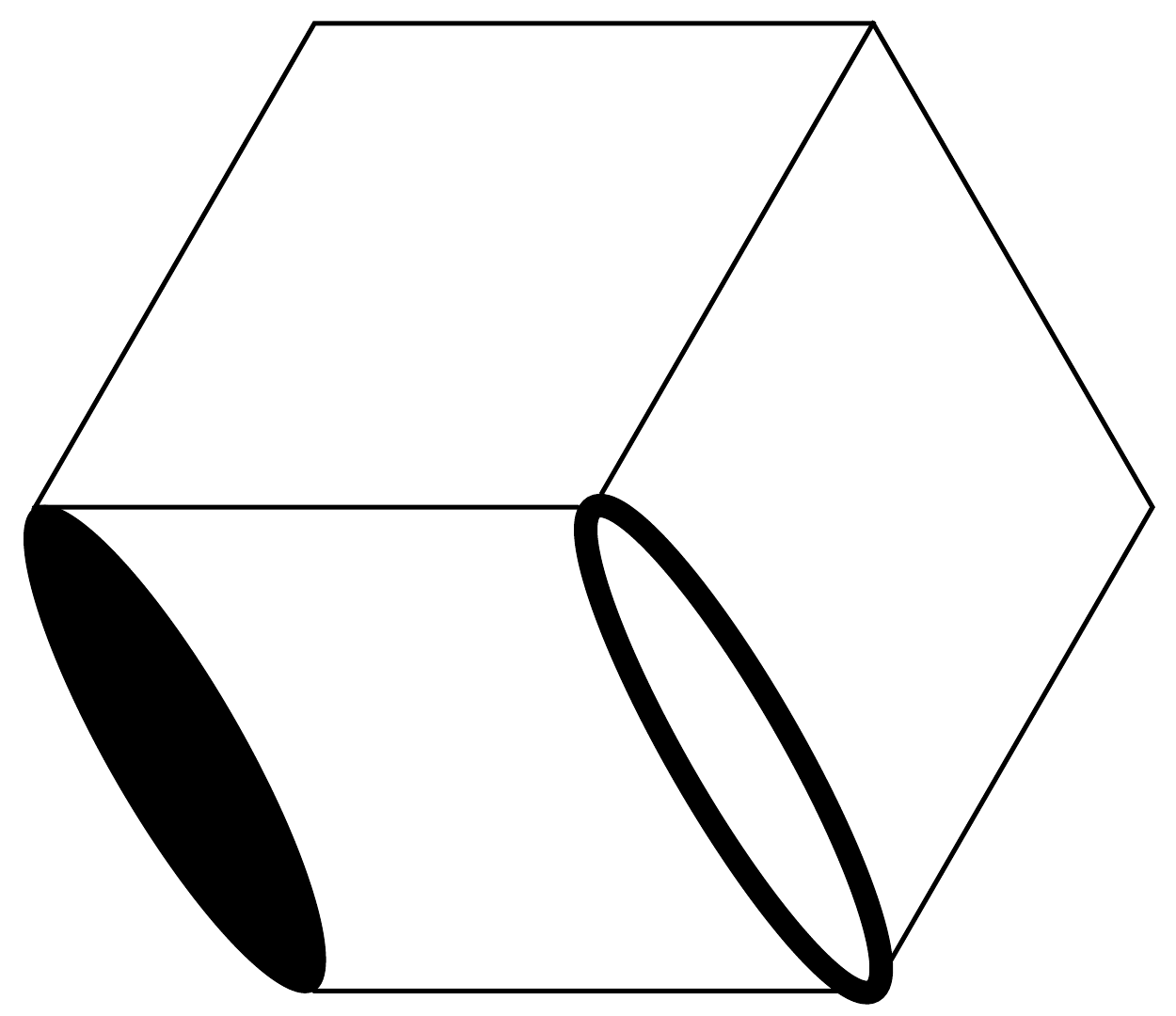}}}}}
\newcommand{\elemC}{\mathord{\vcenter{\hbox{\includegraphics[height=6ex]{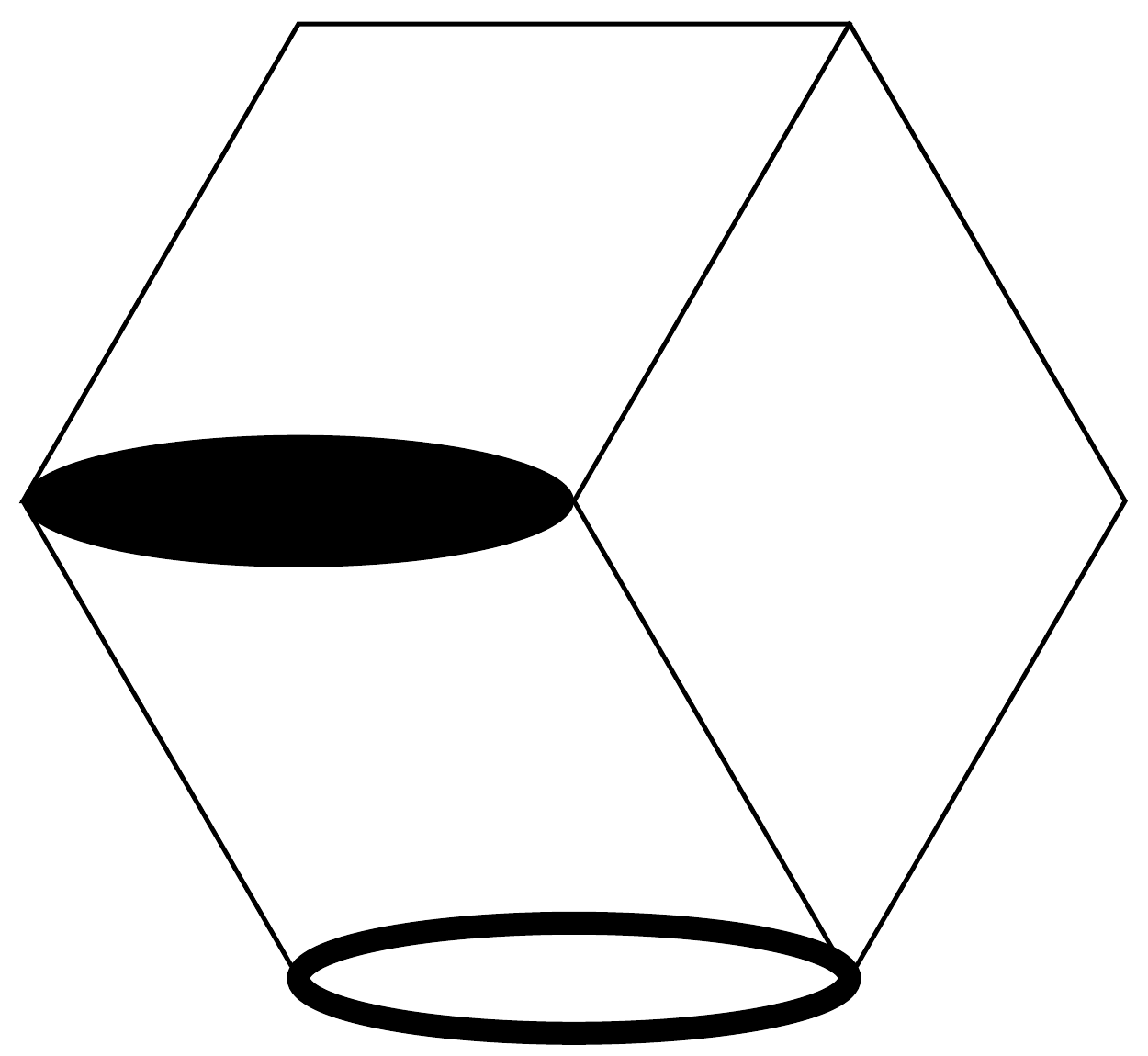}}}}}
\newcommand{\elemBext}{\mathord{\vcenter{\hbox{\includegraphics[height=12ex]{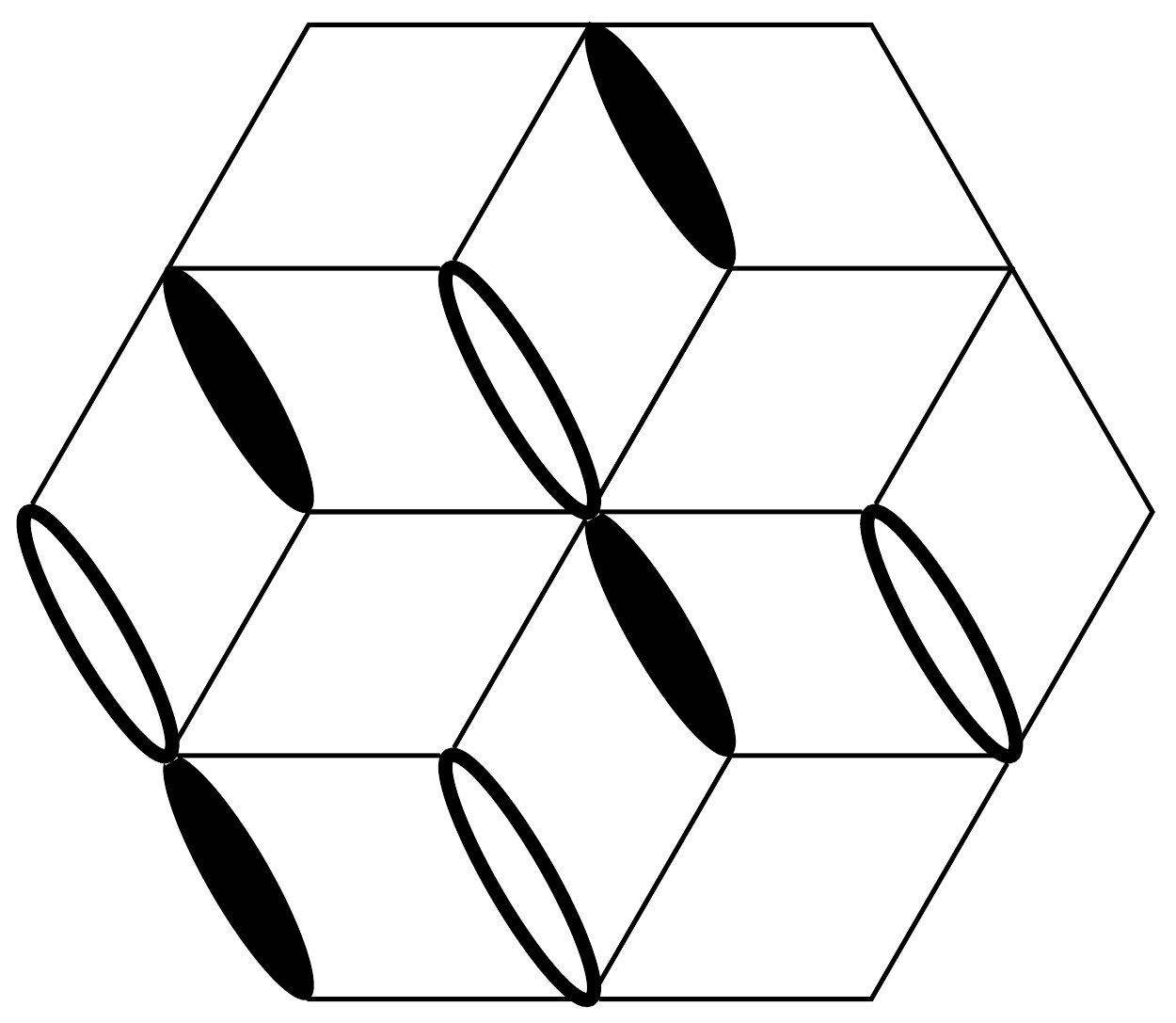}}}}}

\begin{document}
%\preprint{arXiv:1105.xxxx}

\title{Optical conductivity of visons in $Z_2$ spin liquids\\ close to a VBS transition on the kagome lattice}

\author{Yejin Huh}
\affiliation{Department of Physics, Harvard University, Cambridge MA
02138}

\author{Matthias Punk}
\affiliation{Department of Physics, Harvard University, Cambridge MA
02138}

\author{Subir Sachdev}
\affiliation{Department of Physics, Harvard University, Cambridge MA
02138}

\date{\today }

\begin{abstract}
We consider $Z_2$ spin liquids on the kagome lattice on the verge of 
a valence bond solid (VBS) transition, where vortex excitations carrying $Z_2$ magnetic flux -- so-called visons -- condense. 
We show that these vison excitations can couple directly to the external electromagnetic field, even though they carry neither spin nor charge. This is possible via a magneto-elastic coupling mechanism recently identified. \cite{Potter, Hao} For the case of transitions to a 36-site unit cell VBS state the corresponding finite ac-conductivity has a specific power law frequency dependence, which is related to the crossover exponent of the quantum critical point. The visons' contribution to the optical conductivity at transitions to VBS states with a 12-site unit cell vanishes, however.
\end{abstract}


\maketitle

\section{Introduction}

Spin liquids,\cite{Subir, Balents} albeit being Mott insulators, exhibit a finite ac conductivity below the Mott gap provided they have gapless excitations which couple to the external electromagnetic field.\cite{Ioffe, Lee, Bulaevskii, Katsura, Hao} Recent measurements on the kagome material ZnCu$_3$(OH)$_6$Cl$_2$ (Herbertsmithite), which is a strong contender for exhibiting a spin liquid ground-state, indeed showed a specific power law frequency dependence of the ac conductivity below the Mott gap.\cite{Pilon} The nature of the ground-state in this material remains an open issue, however. While experiments are in favor of an almost gapless spin-liquid,\cite{Helton, Han} numerical results of the antiferromagnetic Heisenberg model on the kagome lattice are still controversial. Projected fermion wavefunction studies suggest a U(1) Dirac spin liquid as ground-state\cite{Iqbal}, whereas density matrix renormalization group (DMRG) approaches provide substantial evidence for a gapped $Z_2$ spin-liquid.\cite{White,Balents2,Schollwoeck}. Ref.~\onlinecite{White} in particular indicate that the ground state is close to a phase transition to a 12-site unit cell valence bond solid (VBS). Projected boson wavefunction studies also support a $Z_2$ spin liquid \cite{Motrunich} while large scale exact diagonalization studies remain inconclusive\cite{Laeuchli, Nakano}. Numerical studies of the quantum dimer model imply a 36-site unit cell VBS ground state, which is close to a $Z_2$ spin liquid, separated by a quantum critical point. \cite{Poilblanc}

In a recent paper Potter {\em et al.\/}~\cite{Potter} 
identified three mechanisms which give rise to a finite optical conductivity $\sigma(\omega)$ of gapless spin liquids on the kagome lattice below the Mott gap, all leading to a characteristic $\sigma(\omega) \sim \omega^2$ frequency dependence. They argued that a magneto-elastic coupling, where an applied electric field distorts the lattice and thereby modulates the magnetic exchange couplings, directly couples the external- to the emergent gauge field of a U(1) Dirac spin liquid and gives the largest contribution to the optical conductivity out of the three mechanisms they found. 
Following their approach we show that an external field can also couple to vison excitations of a $Z_2$ spin liquid by the same magneto-elastic coupling mechanism. Indeed, since the visons are $Z_2$-vortices living on the dual lattice, their hopping amplitudes are modulated in accordance with the field-induced distortion of the direct lattice. This mechanism provides a \emph{direct} coupling between the external electromagnetic field and vison excitations of a $Z_2$ spin liquid, which carry neither spin nor charge. 

In the following, motivated by the above mentioned DMRG studies, we focus on the situation of a $Z_2$ spin liquid \cite{sskagome} close to a VBS transition, where the vison gap vanishes. 
In a previous work\cite{Huh} we derived low energy field theories for different $Z_2$ spin liquid to VBS transitions on the kagome lattice. Based on these results we show that the magneto-elastic coupling of visons to the external field leads to a power law frequency dependence of the conductivity for transitions to a VBS state with a 36-site unit cell. This power law comes with a considerably smaller exponent than the $\sim \omega^2$ behavior found in Ref.~\onlinecite{Potter}. 
By contrast, we show that the contribution of visons to the ac-conductivity at transitions to a VBS state with a 12-site unit cell vanishes on symmetry grounds.

Quite generally, a uniform external electric field can only couple to a  time-reversal even spin singlet operator which transforms as a vector under the lattice symmetry group. Physically, this vector operator represents the lattice polarization $\mathbf{P}$, which couples linearly to the external electric field $\mathbf{E}$:
\begin{equation}
\delta H \sim \mathbf{P} \cdot \mathbf{E}
\end{equation}
Since visons transform projectively under lattice symmetry operations, we have to find an operator in terms of the vison fields which transforms as a vector under the corresponding projective symmetry group (PSG).\cite{Wen} In our previous paper\cite{Huh} we showed  explicitly how the visons transform under the PSG and we are going to utilize these results in order to construct such a vector operator on symmetry grounds. Moreover, we show that this operator is indeed the unique polarization operator which couples to the external field by the above mentioned magneto-elastic mechanism.

\section{Magneto-elastic coupling between visons and the external electromagnetic field}

The basic idea behind the magneto-elastic coupling mechanism of Ref.~\onlinecite{Potter} is the fact that positively charged copper ions can be displaced within the unit cell by an applied uniform external electric field, which in turn leads to a modulation of the super-exchange amplitudes between neighboring spins. The corresponding perturbation to the Heisenberg model takes the form 
\begin{equation}
\delta H = \sum_{\langle i, j \rangle} \delta \tilde{J}_{ij} \, \S_i \cdot \S_j \ ,
\end{equation}
where $\S_i$ describes a spin-1/2 operator on lattice site $i$. The pattern of modulated super-exchange amplitudes $\delta  \tilde{J}_{ij}$ which couples to the external field can be inferred on symmetry grounds, as shown explicitly in Ref.~\onlinecite{Potter}.

\begin{figure}
\begin{center}
\includegraphics[width=0.76\columnwidth]{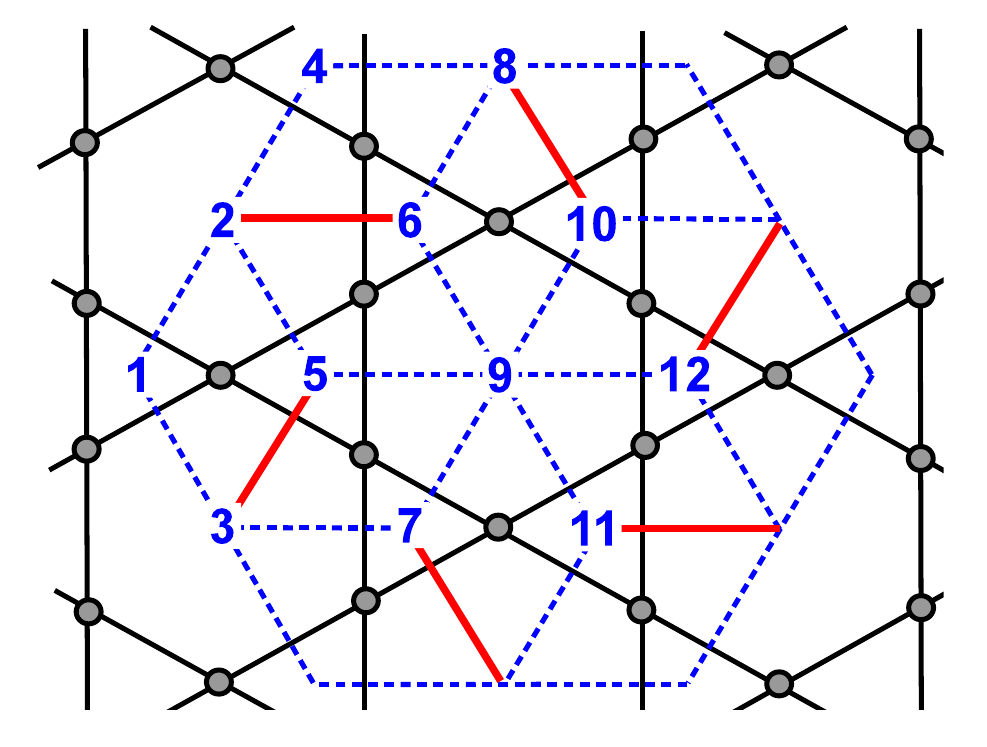}
\caption{(Color online) kagome lattice (black solid lines) and its dual dice lattice (blue dashed lines). Shown is one extended 12-site unit cell of the dice lattice together with a gauge choice of frustrated bonds ($J_{ij} = -J$ shown as solid red lines; other dice lattice bonds have $J_{ij}=J$) for the effective frustrated Ising model describing the vison excitations (see text). 
%The elementary dice lattice unit cell has three sites.
}
\label{fig1}
\end{center}
\end{figure}

For our purpose, the important low energy excitations of a $Z_2$ spin liquid close to a VBS transition are vortices carrying $Z_2$ magnetic flux, so-called visons. These are described by an effective fully frustrated transverse field Ising model on the dual dice lattice (see Fig.~\ref{fig1}). Its low energy properties are captured by a field-theory of soft-spin modes\cite{Huh} with a Lagrangian of the form ($\tau$ denotes the imaginary time)
\begin{equation}
\mathcal{L}_v =  \sum_{i \leq j} \phi_i(\tau) \big[ (-\partial_\tau^2 +m^2) \delta_{i,j} - J_{ij} \big]  \phi_j(\tau) + \dots
\label{lagrangian_v}
\end{equation}
together with the full frustration condition
\begin{equation}
\prod_\text{plaq.} \text{sign}(J_{ij}) = -1 \ ,
\end{equation}
where the product is over an elementary plaquette of the dice lattice.
A specific gauge choice for the $J_{ij}$'s which satisfies this frustration condition is shown in Fig.~\ref{fig1}. Note that any gauge choice requires an extended unit cell. In Eq.~\eqref{lagrangian_v} we didn't explicitly include higher order terms in the fields $\phi$ that are allowed by symmetry and describe interactions between the visons. Furthermore, note that the hopping amplitudes $J_{ij}$ are not equal to super-exchange amplitudes $\tilde{J}_{ij}$ in the Heisenberg model above, but they are expected to be of the same order, since the magnetic super-exchange coupling is the only energy scale in the problem.

Important for our considerations is the fact that a distortion of the kagome lattice inevitably leads to a distortion of the dual dice lattice. This implies that the hopping amplitudes $J_{ij}$ of the visons on the dice lattice are modulated by the magneto-elastic coupling in a way similar to the modulation of the exchange amplitudes in the Heisenberg model, giving rise to a perturbation of the form
\begin{equation}
\mathcal{L}_\text{m-e} =  - \sum_{i<j} \delta_{ij}  \, J_{ij}  \phi_i(\tau) \phi_j(\tau)  =  \sum_\ell \mathbf{P}(\ell) \cdot \mathbf{E} \ ,
\end{equation} 
where $\mathbf{P}(\ell)$ is the polarization of an elementary unit cell of the dice lattice and the sum on $\ell$ runs over unit cells. We have $\delta_{ij}>0$ ($\delta_{ij}<0$) for a squeezed (stretched) bond, where the hopping amplitude increases (decreases). 
The absolute value of $\delta_{ij}$, which parametrizes the change of the vison hopping amplitude with respect to the applied electric field $E$ can be simply estimated as
\begin{equation}
|\delta_{ij}|  \sim \frac{e E}{K_\text{Cu} a}
\end{equation} 
with $K_\text{Cu}$, $e$, and $a$ as effective spring constant of the copper ions, elementary charge, and lattice constant, respectively, and $E =|\mathbf{E}|$ is the external electric field. The pattern of modulated hopping amplitudes can be inferred on symmetry grounds. Following Ref.~\onlinecite{Potter} we construct an irreducible representation of the modulated bonds $\delta_{ij}  J_{ij} \phi_i \phi_j$  which transforms as a vector under the symmetry group of the dice lattice.\cite{footnote1} This operator is proportional to the lattice polarization operator $\mathbf{P}(\ell)=(P_x,P_y)$ which takes the form
\begin{eqnarray}
P_x &=& \frac{e a}{J} \left[ \elemA + \elemB + 2 \ \elemC \right] \label{Px_dice} \\
P_y &=& \frac{e a}{J} \, \sqrt{3} \left[ \elemA - \elemB  \right]  \ , \label{Py_dice}
\end{eqnarray}
where black (white) ellipses correspond to squeezed bonds $+\delta J_{ij} \phi_i \phi_j$ (stretched bonds $-\delta  J_{ij} \phi_i \phi_j$) within one elementary unit cell of the dice lattice and we have defined
\begin{equation}
\delta = \frac{J}{K_\text{Cu} a^2}
\end{equation}
with $J$ denoting the absolute value of the nearest neighbor vison hopping amplitude. The dimensionless parameter $\delta$ represents the magneto-elastic coupling strength and is typically much smaller than one. While the elastic energy $K_\text{Cu} a^2$ is on the order of 1eV, $J$ takes values on the order of 10 meV.
On the extended 12-cite unit cell shown in Fig.~\ref{fig1} the basis elements take the explicit form
\begin{eqnarray}
\elemB &\equiv& \elemBext  \label{dice_extended} \\
&=& \delta  J \,  \big[ \phi_2(\R) \phi_5(\R)- \phi_1(\R) \phi_3(\R) \notag \\
&&+ \phi_3(\R) \phi_4(\R+2 \v -2 \u) - \phi_8(\R) \phi_{10}(\R) \notag \\
&&- \phi_6(\R)\phi_9(\R)   + \phi_7(\R) \phi_8(\R+2 \v-2 \u) \notag \\
&& + \phi_9(\R) \phi_{11}(\R) - \phi_{12}(\R) \phi_2(\R+2 \v) \big] \notag
\end{eqnarray}
and similar expressions for the two other basis elements. Note that the sign of the gauge choice from Fig.~\ref{fig1} has been taken into account in the equation above.
$\mathbf{R}=2 m \, \u +2 n \, \v$ denotes the lattice vector of the extended 12-site unit cell, with $m,n\in \mathbb{Z}$ and $\u=(3/2,\sqrt{3}/2)$, $\v=(3/2,-\sqrt{3}/2)$ as the two basis vectors of the dice lattice.

Also note that the specific hopping amplitudes beyond nearest neighbors, which had to be introduced in Ref.~\onlinecite{Huh} in order to give a dispersion to the otherwise flat vison bands, are not modulated here, because the distance between these sites doesn't change.

\section{Optical conductivity}

\subsection{Transition to a 12-site VBS state}

We start by briefly reviewing the necessary results of Ref.~\onlinecite{Huh}.
The field theory for visons at the transition from a $Z_2$ spin liquid to a VBS state with a 12-site unit cell is an O(4) theory with additional 4th order terms which break the symmetry down to $GL(2,\mathbb{Z}_3)$. Its Lagrangian takes the form
\begin{eqnarray}
\mathcal{L} &=& \sum_{n=1...4}((\nabla \psi_n)^2 + (\partial_\tau \psi_n)^2 + r \psi_n^2 + u \psi_n^4) \notag \\
&&+ a \sum_{n < m} \psi_n^2 \psi_m^2   + b \big[ \psi_1^2 (\psi_2 \psi_3-\psi_2 \psi_4+\psi_3 \psi_4)  \notag \\
&&+ \psi_2^2 (\psi_1 \psi_3+\psi_1 \psi_4-\psi_3 \psi_4) \notag \\
&&+ \psi_3^2 (\psi_1 \psi_2-\psi_1 \psi_4+\psi_2 \psi_4) \notag \\
&& - \psi_4^2 (\psi_1 \psi_2+\psi_1 \psi_3+\psi_2 \psi_3) \big]. \label{GL1}
\end{eqnarray}
The fields $\psi_i$ are linear combinations of the soft-spin fields $\phi_i$ and correspond to the modes which become critical at the VBS transition, i.e.
\begin{equation}
\phi_j(\R) = \sum_{n=1\dots4} \psi_n v_j^{(n)} \ ,
\label{modeexpansion}
\end{equation}
where $v_j^{(n)}$ denotes the four eigenvectors corresponding to the highest, four-fold degenerate eigenvalue of the Fourier transform of the hopping matrix $J^{(ij)}_{\q=0}$ at momentum $\q=0$ and $j=1,\dots,12$ is a sublattice index in the extended 12-site unit cell (see Fig.~\ref{fig1} and Ref.~\onlinecite{Huh} for details). Note that in this case $\phi_j(\mathbf{R})$ does not depend on the lattice vector $\mathbf{R}$ of the extended 12-site unit cell.

The four fields $(\psi_1, \psi_2, \psi_3, \psi_4)$ transform under lattice symmetries via a four dimensional representation of the PSG generated by lattice translations $T_\u$, six-fold rotations $R_6$ and reflections about the x-axis $I_x$, which take the form 
\begin{eqnarray}
T_\u &=& \begin{bmatrix}
0 & 0 & -1 & 0 \\
0 & 0 & 0 & -1 \\
1 & 0 & 0 & 0 \\
0 & 1 & 0 & 0
\end{bmatrix}, \
I_x = \begin{bmatrix}
0 & 1 & 0 & 0 \\
1 & 0 & 0 & 0 \\
0 & 0 & 1 & 0 \\
0 & 0 & 0 & -1
\end{bmatrix}, \notag \\
R_6 &=& \begin{bmatrix}
0 & 0 & 1 & 0 \\
1 & 0 & 0 & 0 \\
0 & 1 & 0 & 0 \\
0 & 0 & 0 & 1
\end{bmatrix} \ .
\label{PSG12}
\end{eqnarray}
We are looking for an operator which transforms as a vector under \eqref{PSG12}. Such an operator which couples to lattice distortions (i.e. terms in the vison Hamiltonian of the form $\delta  J_{ij} \phi_i \phi_j$) can only be bilinear in the fields $\psi$. Moreover, such an operator has to change its sign under rotations by $180$ degrees if it is supposed to transform like a vector. Now one can readily show from \eqref{PSG12} that $(R_6)^3 = \mathbf{1}$, i.e. the fields transform back to themselves after 180 degree rotations, thus it is not possible to construct a vector operator out of the fields $\psi_i$ without invoking spatial gradients. Since the external electric field is homogeneous and only couples to the zero momentum component of such a vector operator, any PSG invariant bilinear term in the $\psi_i$'s involving gradients is ruled out. In principle it would be possible to construct a fourth order term involving gradients which transforms as a vector and gives a non-zero contribution to the conductivity when coupled to a homogeneous external field, but such an operator cannot correspond to the lattice polarization operator as it is not a bilinear. Since we are not aware of a mechanism which couples the electric field to a quadrilinear operator in the fields $\psi$, we do not pursue this route further at the moment.

We can also directly show that the lattice polarization operator vanishes for the case of a transition to a 12-site VBS state by using Eqs.~\eqref{Px_dice}, \eqref{Py_dice}, \eqref{dice_extended} and \eqref{modeexpansion}. Since $\phi_j(\R)$ doesn't depend on the lattice vector $\R$ we get
\begin{eqnarray}
\elemB &=& \delta  J \sum_{n,m=1}^4 \psi_m \psi_n  \Big[ v^{(n)}_2 v^{(m)}_5- v^{(n)}_1 v^{(m)}_3 \notag \\
&&- v^{(n)}_8 v^{(m)}_{10}- v^{(n)}_6 v^{(m)}_9 + v^{(n)}_9 v^{(m)}_{11} \notag \\
&&+ v^{(n)}_7 v^{(m)}_8 + v^{(n)}_3 v^{(m)}_4- v^{(n)}_{12} v^{(m)}_2 \Big] \nonumber\\
&=& 0 \ , 
\end{eqnarray}
where the last equation follows after using the explicit form of the eigenvectors $v_j^{(n)}$ from Ref.~\onlinecite{Huh}.
Similarly also the other two basis elements that are used to construct the polarization operator shown in Eqs.~\eqref{Px_dice} and \eqref{Py_dice} vanish. Consequently it is not possible to construct a lattice polarization operator which couples to the external electric field via the magneto-elastic mechanism. In fact we see that the bonds related by a 180 degree rotation cancel each other, confirming our earlier argument.

\subsection{Transition to a 36-site VBS state}

For a transition to a VBS state with a 36-site unit cell  the visons are described by an O(8) theory with additional fourth and sixth order terms which break the symmetry down to $GL(2,\mathbb{Z}_3) \times D_3$. The explicit form of the Lagrangian and the corresponding eight-dimensional matrix representation of the PSG can be found in Ref.~\onlinecite{Huh}. In contrast to the previous case, we can straightforwardly construct a homogeneous bilinear operator which transforms like a vector under the PSG by making the ansatz ($\bm{\mathcal{P}} =(\mathcal{P}_x,\mathcal{P}_y)$)
\begin{equation}
\mathcal{P}_x=\sum_{i,j=1\dots8} a_{ij} \psi_i \psi_j, \hspace{0.25cm} \mathcal{P}_y=\sum_{i,j=1\dots8} b_{ij} \psi_i \psi_j
\end{equation}
and determining the coefficients of the matrices $(\mathbf{a})_{ij} \equiv a_{ij}$ and $(\mathbf{b})_{ij} \equiv b_{ij}$ via
\begin{eqnarray}
 R_6^{\mathsf{T}} \, \mathbf{a} \, R_6  &=& (\mathbf{a}+\sqrt{3}\, \mathbf{b})/2 \\
 R_6^{\mathsf{T}} \, \mathbf{b} \, R_6  &=& (\mathbf{b}-\sqrt{3} \, \mathbf{a})/2 \\
 T_\u^{\mathsf{T}} \, \mathbf{a} \, T_\u  &=& \mathbf{a} \\
 T_\u^{\mathsf{T}} \, \mathbf{b} \, T_\u  &=& \mathbf{b} \\
 I_x^{\mathsf{T}} \, \mathbf{a} \, I_x  &=& \mathbf{a}\\
 I_x^{\mathsf{T}} \, \mathbf{b} \, I_x   &=& -\mathbf{b} \ .
\end{eqnarray}
This vector operator is uniquely determined up to a constant prefactor and takes the form 
\begin{eqnarray}
\mathcal{P}_x &\sim& \Big[ e^{i 5 \pi/6} \big( 2 \psi_1 \psi_2^*+\psi_1\psi_3^*+\psi_1^* \psi_4+2 \psi_3 \psi^*_4- \psi_2 \psi_4^* \big) \notag \\
&& +i \psi_2 \psi_3^* \Big] \ + \ \text{c.c.} \label{Px} \\
\mathcal{P}_y &\sim& \sqrt{3} \Big[ e^{i 5 \pi/6} \big( \psi_1^* \psi_4-\psi_1 \psi_3^*+\psi_2 \psi_4^* \big) + i \psi_2 \psi_3^* \Big] \ \notag \\
&&+ \ \text{c.c.}
\label{Py}
\end{eqnarray} 
As we will show now, this operator corresponds to the unique lattice polarization operator $\mathbf{P} = e a \delta \, \bm{\mathcal{P}}$ in Eqs.~\eqref{Px_dice} and \eqref{Py_dice}. 
For transitions to a VBS state with a 36-site unit cell, the soft-spin modes $\phi$ are related to the critical modes $\psi$ via\cite{Huh}
\begin{equation}
\phi_j(\R) =  e^{i \Q_1 \cdot \R} \sum_{n=1...4} \psi_n v^{(n)}_{\Q_1,  j} + \text{c.c.}
\label{phi_36}
\end{equation}
where  $\q = \pm \mathbf{Q}_1 =  \big( 0, \ \pm \frac{2 \pi}{3 \sqrt{3}} \big)$ are the momenta where the vison dispersion has its minima. When inserting Eq. \eqref{phi_36} into Eqs.~\eqref{Px_dice} and \eqref{Py_dice} it is important to keep only the zero-momentum components (i.e.~the terms independent of the lattice vector $\R$) since the external electric field is homogeneous and will only couple to such terms. Using the shorthand notation
\begin{equation}
\mu_j = \sum_{n=1...4} \psi_n v^{(n)}_{\Q_1,  j} \\
\end{equation} 
we obtain
\begin{eqnarray}
\elemB &=& \delta J \big( \mu_2 \mu_5^*- \mu_1 \mu_3^*- \mu_8 \mu_{10}^*- \mu_6 \mu_9^*+ \mu_9 \mu_{11}^* \notag \\ 
&& \ \ + \mu_7 \mu_8^* e^{-i2\Q_1\cdot(\v-\u)} + \mu_3 \mu_4^* e^{-i2\Q_1\cdot(\v-\u)} \notag \\
&& \ \ - \mu_{12} \mu_2^* e^{-i2\Q_1\cdot\v}) +\text{c.c.} \nonumber\\
&=& \frac{\delta J}{\sqrt{2}} \big(-e^{-i \pi/6}\psi_1\psi_2^*+ e^{i \pi/6}\psi_1\psi_4^* \notag \\
&& \ \ -e^{-i \pi/6}\psi_3\psi_4^* -i\psi_2\psi_3^*\big) + \text{c.c.}
\end{eqnarray}
The exponential factors arise from bonds connecting sites in adjacent unit cells. Using this and similar expressions for the two other basis elements, we reproduce Eqs. \eqref{Px} and \eqref{Py} exactly up to a multiplicative constant. 

The frequency dependence of the conductivity can now be obtained using Kubo's formula
\begin{equation}
\sigma(\omega) \sim  \omega \langle \mathbf{P}_\omega \mathbf{P}_{-\omega} \rangle \ ,
\end{equation}
where the frequency dependence of the polarization correlation function is determined by the scaling dimension of the polarization operator $\mathbf{P}_\omega$. In Ref.~\onlinecite{Huh} we couldn't find a stable fixed point at one-loop order for our O(8) theory including the additional fourth order terms. Since the O(8) theory has a small anomalous dimension, we do not expect the scaling dimensions of the true critical theory to be significantly different from that of the O(8) model. 
Thus making the simplifying assumption that the critical point between the $Z_2$ spin-liquid and the VBS phase is described by the standard O(8) Wilson-Fisher fixed point, the polarization operator in Eqs.~\eqref{Px} and \eqref{Py} can be expanded in terms of components of the traceless symmetric bilinear tensor operator of the O(8) model
\begin{equation}
T_{mn}(x) = \psi_n(x) \psi_m(x) -  \frac{\delta_{m,n}}{8} \sum_k \psi_k(x)^2 \ ,
\end{equation}
where $\psi_n$ ($n \in \left\{1, \dots, 8 \right\} $) now represents the real- and imaginary parts of the four complex fields in Eqs.~\eqref{Px} and \eqref{Py}. 
We denote the scaling dimensions of this tensor operator, and thus the polarization operator, by
$\Delta_{T}$.
Then the frequency scaling of the ac conductivity in $d=2$ dimensions is given by
\begin{equation}
\sigma(\omega) \sim \omega^{2 \Delta_T - 2} \ .
\label{endresult}
\end{equation}
The scaling dimension can be expressed in terms of the crossover exponent $\phi_T$ by $\Delta_T = 3-\phi_T/\nu$, and has been evaluated to six loops\cite{Calabrese} as $\phi_T  = 1.55$.
Using $\nu = 0.830$ \cite{Sobolev} we obtain
\begin{equation}
2 \Delta_T - 2 = 0.27
\end{equation}
This exponent is considerably smaller than two, giving rise to a large optical response compared to the mechanisms discussed in Ref.~\onlinecite{Potter} at frequencies below the magnetic super-exchange coupling. 
The resulting expression for the frequency dependent conductivity finally takes the form
\begin{equation}
\sigma(\omega) \approx \frac{e^2}{h} \Big( \frac{J}{K_\text{Cu} a^2} \Big)^2 \Big(\frac{\omega}{J}\Big)^{0.27}
\end{equation} 
where we have reintroduced Planck's constant $h$. Note again that $J$ denotes the nearest neighbor vison hopping amplitude here, which is expected to be on the order of the magnetic super-exchange coupling of the underlying Heisenberg model, since this is the only energy scale in the problem.   

\section{Discussion}

We calculated the optical conductivity of $Z_2$ spin liquids at a quantum critical point to a VBS state with a 36-site unit cell, 
and found a power-law frequency dependence of the conductivity $\sigma(\omega) \sim  \omega^{0.27}$, a particularly small exponent. This indicates a large optical conductivity at low frequencies at the critical point, and the enhancement should
also persist away from the critical point.

This calculation is based on the simplifying assumption that the transition is described by the Wilson-Fisher fixed point of the O(8) model. It is still an open question, if the full theory including the O(8) symmetry breaking terms presented in Ref.~\onlinecite{Huh} exhibits a stable fixed point, and it would be interesting to study its critical behavior in the future. 
Going beyond a one-loop calculation is a daunting task, however, given the complicated form of the Lagrangian (see Eq.~(3.24) in Ref.~\onlinecite{Huh}). A viable approach would be to study the frustrated Ising model directly using Monte-Carlo simulations.
In any case, the smallness of the optical conductivity exponent relies mainly on the fact that polarization operator is a bilinear in the
field of the O(8) model, with no additional spatial gradients, and so we can reasonably expect a small exponent at any possible
fixed point which breaks the O(8) symmetry down to $GL(2,\mathbb{Z}_3) \times D_3$.

\acknowledgements 

We gratefully acknowledge very helpful discussions with Andrew Potter. This research was supported by U.S. NSF under grant DMR-1103860, and by the U.S. Army Research Office Award W911NF-12-1-0227.


\begin{thebibliography}{999}
\bibitem{footnote1}Note that only the combination $J_{ij} \phi_i \phi_j$ is gauge invariant. This is why we construct an irreducible vector representation of the modulated bond strengths $J_{ij} \phi_i \phi_j$, and not of the modulated hopping amplitudes alone.
\end{thebibliography}
\end{document}